\documentclass[floatfix,showkeys,preprint,citeautoscript,showpacs,amsmath,amssymb,epsf,pra]{revtex4}
\usepackage{graphicx}
\usepackage{dcolumn}
\usepackage{color}
\usepackage{float}
\usepackage{amsmath}
\usepackage{soul}
\usepackage[section]{placeins}
\usepackage{booktabs}
\usepackage{multirow}

\begin{document}

\title{Study of self-interaction errors in density functional predictions of dipole polarizabilities and ionization energies of water clusters using Perdew-Zunger and locally scaled self-interaction corrected methods}
\author{Sharmin Akter$^1$, Yoh Yamamoto$^2$, Carlos M. Diaz$^1$, Koblar A. Jackson$^3$, Rajendra R. Zope$^2$, and Tunna Baruah$^2$}
\affiliation{$^1$Computational Science Program, The University of Texas at El Paso, El Paso, Texas, 79968}
\affiliation{$^2$Department of Physics, University of Texas at El Paso, TX, 79968}
\affiliation{$^3$Physics Department and Science of Advanced Materials Program, Central Michigan University, Mt. Pleasant Michigan, 48859}

\date{\today}
\begin{abstract}
   We studied the effect of self-interaction error (SIE) on the static dipole 
   polarizabilities of water clusters modelled with three increasingly sophisticated, non-empirical density functional
   approximations (DFAs), \textit{viz.} the local spin density approximation (LDA), the 
   Perdew-Burke-Ernzherof (PBE) generalized-gradient approximation (GGA), and the strongly constrained and
   appropriately normed (SCAN) meta-GGA, using the Perdew-Zunger self-interaction-correction (PZ-SIC) energy functional 
   in the Fermi-L\"owdin orbital SIC (FLO-SIC) framework.  Our results show that while all three DFAs 
   overestimate the cluster polarizabilities, the description systematically
   improves from LDA to PBE to SCAN. The self-correlation free SCAN
   predicts polarizabilities
   quite accurately with a mean absolute error (MAE) of 0.58 Bohr$^3$ with respect to  coupled cluster singles and doubles (CCSD) values.
   Removing SIE using PZ-SIC correctly reduces the DFA polarizabilities, but over-corrects, 
   resulting in underestimated polarizabilities in SIC-LDA, -PBE, and -SCAN.
   Finally, we applied a recently proposed
   local-scaling SIC (LSIC) method using a {\em quasi} self-consistent scheme and using the kinetic energy density ratio 
   as an iso-orbital indicator. The results  show
   that the LSIC polarizabilities are in excellent agreement with 
    mean absolute error of 0.08 Bohr$^3$ for LSIC-LDA and 0.06 Bohr$^3$ for LSIC-PBE with most recent CCSD polarizabilities.
   Likewise, the ionization energy estimates as an absolute of highest occupied energy eigenvalue predicted by LSIC are also in excellent agreement with 
   CCSD(T) ionization energies with MAE of 0.4 eV for LSIC-LDA and 0.04 eV for LSIC-PBE. The LSIC-LDA predictions of
   ionization energies are comparable to the reported GW ionization energies while the LSIC-PBE 
   ionization energies are more accurate than reported GW results.

\end{abstract}    

\maketitle

\section{Introduction} \label{sec:introduction}
 Water plays a central role in many areas of science.  For that reason, it has been the subject of many computational studies in chemical physics.\cite{ball2008water,fecko2003ultrafast,wernet2004structure,soper2008quantum,skinner2013benchmark,pedroza2018bias,elton2016hydrogen,pedroza2015local,piaggi2020phase,sommers2020raman,Chen10846,santra2015local}  These studies have led to important insights into the atomic-level behavior of water, but they also allow detailed comparisons of the methods used for the calculations.  Generally speaking,  \textit{ab initio} methods at the coupled-cluster level of theory reliably predict the observed properties of water. Density functional theory (DFT) -based methods are more computationally efficient and, with increasing sophistication of the density functional approximation (DFA) used, can give accurate predictions for many properties. 
 Most DFAs, however, are known to over-bind water clusters, particularly at the local density approximations level,\cite{LAASONEN1992,LAASONEN1993,Lee,Car1993} 
 fail to correctly predict the structural energy ordering of cluster isomers,\cite{Santra2008, Truhlar2008} transition pressure for phase transition of ice,\cite{Santra2011} and density in liquid phase.\cite{PhysRevE.68.041505,doi:10.1063/1.1630560,doi:10.1063/1.1782074}
 For a recent perspective on the performance of DFT methods for describing water, see Ref.~\cite{doi:10.1063/1.4944633}.
 The recently developed strongly constrained and appropriately normed (SCAN) meta-GGA\cite{SCAN} further improves the description of water clusters, liquid water, and ice,\cite{Sun2016, Chen10846} but shortcomings such as over-binding tendency remain. 
 %
 We recently showed that much of the over-binding can be eliminated by removing self-interaction error (SIE) from the DFA. SIE is caused by an incomplete cancellation of self-Coulomb and self-exchange energies in the DFA.  The Perdew-Zunger self-interaction correction\cite{doi:10.1063/1.4820448} (PZ-SIC) removes the self-interaction on an orbital by orbital basis.  Using the PZ-SIC energy functional in the Fermi-L\"owdin orbital self-interaction correction (FLO-SIC) method to remove self-interaction, we showed that the overbinding in 
 SCAN\cite{SCAN} is reduced by a factor of 3 to about 10 meV per molecule in small water clusters.\cite{FLOSIC_WATER_PNAS} Furthermore, SCAN is known to predict the correct energetic ordering of the isomers of the water hexamer, a celebrated test for DFAs, and this is not changed in the FLO-SIC-SCAN calculations.

 Because water is a polar solvent, its dielectric properties are important. Several \textit{ab initio} and DFT calculations have examined how the dipole moments and static dipole polarizabilities of (H$_2$O)$_n$ clusters evolve with $n$,  \cite{Jensen,Tu,doi:10.1063/1.1573171,doi:10.1002/qua.20308,C001007C,doi:10.1063/1.2210937} as well as how these properties can be broken into many-body contributions.\cite{Paesani_pol_2013} 
  Hammond \textit{et al.} carried out calculations of the dipole polarizability of water clusters up to $n = 12$ at the coupled-cluster level of theory\cite{Hammond} and using several DFAs, using cluster geometries optimized at the MP2 level.\cite{Xantheas_MP2_1, Xantheas_MP2_2} By carefully studying details such as the effect of basis set and the level of theory, their results provide excellent reference values. 
 Similarly, Alipour and co-workers studied water cluster polarizabilities with coupled-cluster and hybrid and double hybrid DFT methods.\cite{Alipour2013, Alipour2017}  
 While these last two studies shed some light on how DFT-based methods perform for dielectric properties, they do not directly address the effect of SIE.  That is the topic of this paper.  
 
 It is clear that SIE should impact calculated dipoles and polarizabilities.  A well-known effect of electron self-interaction is that the potential “seen” by electrons is too-repulsive in the asymptotic region of a neutral system, decaying exponentially to zero, rather than with the correct $-1/r$ dependence.  The result is that electrons are too loosely bound in a DFA calculation, as can be seen, for example, in orbital energies that are too high in comparison with removal energies.\cite{Jackson_2019} 
 Similarly, in calculations of atomic anions, the highest occupied electronic states have positive orbital energies, implying that the anions are actually unstable.  Removing self-interaction via FLO-SIC calculations restores the correct asymptotic behavior of the potential and improves the description of related properties.  In a recent study of water cluster anions,\cite{C9CP06106A} for example, we showed that FLO-SIC significantly improves the prediction of vertical detachment energies (VDEs).  For clusters up to the hexamer, the orbital energy for the highest state in FLO-SIC-PBE predicted the corresponding VDE with a mean average error of only 17 meV compared to CCSD(T) values.  In the context of molecular dipoles, the effect of self-interaction is to destabilize the anionic part of a molecule relative to the cationic part, resulting in too little charge transfer and dipoles that are too small.  Removing self-interaction from a large test set of molecules improved the calculated dipole moments, though the corrections tended to be too large, yielding dipoles that are somewhat too large compared to reference values.  On the other hand, electrons that are too loosely bound respond too strongly to an applied electric field, leading to polarizabilities that are too large.  In a study of neutral and singly charged atoms,\cite{PhysRevA.100.012505} removing self-interaction generally reduces the predicted polarizabilities, improving agreement with experimental values.  But in this case, too, FLO-SIC tends to over-correct, resulting in predictions that underestimate experiment.

Here we report the results of using FLO-SIC-DFA methods to compute the dipole moments and polarizabilities for (H$_2$O)$_n$ clusters up to $n = 20$.  We have used three increasingly sophisticated,\cite{Perdew-Schmidt} non-empirical DFAs for our study --- the LDA in the Perdew-Wang version, the Perdew, Burke, and Ernzerhof (PBE) GGA,\cite{PBE} and SCAN.\cite{SCAN}  We find that removing SIE increases calculated dipole moments and decreases polarizabilities.  We also show that the FLO-SIC corrections overshoot the relevant reference values.  A second objective of this paper is therefore to show the performance of a recent local-scaling self-interaction correction\cite{doi:10.1063/1.5129533} (LSIC) method for the prediction of dipoles and polarizabilities.  LSIC uses a position-dependent factor based on the local character of the charge density to scale down the energy densities of the SIC.  We show that a quasi-self-consistent implementation of LSIC reduces the over-correction of polarizabilities and yields calculated values that are in the best agreement of all the DFT-based methods with reference values.

 
 
 Some details about the FLO-SIC and LSIC methods is provided in Sec.~\ref{sec:computationaldetails}. The results and discussion are presented in Sec.~\ref{sec:results}. In Sec.~\ref{sec:summary} we report our summary.

 

  \section{Methods and Computational Details} \label{sec:computationaldetails}
 
 The Fermi-L\"owdin orbital-based self-interaction correction (FLO-SIC) method\cite{FERMI1,FERMI2,PhysRevA.95.052505,AAMOP} is used for the calculations presented here. It is implemented in the FLO-SIC code\cite{FLOSIC_code} which is based on the UTEP version of the NRLMOL code. FLO-SIC retains the principal features of NRLMOL, including optimized Gaussian basis functions\cite{NRLMOL_basis} and a highly accurate numerical integration grid scheme.\cite{NRLMOL_mesh} 
 
 FLO-SIC makes use of the orbital-dependent PZ-SIC energy functional. Unlike for DFAs, where the energy depends only on the total density and any unitary transformation of the orbital wave functions yields the same energy, minimizing the PZ-SIC total energy requires optimizing both the total density and the choice of orbitals used to evaluate the energy.  In FLO-SIC, localized Fermi-L\"owdin orbitals (FLOs) are used.\cite{FERMI1}  Fermi orbitals are first constructed as follows,
 \begin{equation}\label{eq:FO}
 \phi^{FO}_{i,\sigma}(\vec{r})=\frac{\sum_j \psi_{j,\sigma}(\vec{a}_i)\psi_{j,\sigma}(\vec{r})}{\sqrt{\rho({\vec{a}_i})}} .
 \end{equation}
 Here $i$ and $j$ are the orbital indices, $\sigma$ is the spin index, $\rho$ is the electron density, and the $\psi_{j,\sigma}$ are any set of orbitals that span the occupied space.  In practice, the canonical orbitals are typically used.  $\vec{a}_i$ is a position in space called a Fermi orbital descriptor (FOD). One such FOD is required for each Fermi orbital.  The Fermi orbitals resulting from Eq.~(\ref{eq:FO}) are normalized but not mutually orthogonal. They are orthogonalized using the L\"owdin symmetric orthogonalization scheme.\cite{lowdin1950non}
 
 As shown in Ref.~\cite{FERMI2,AAMOP}, the derivatives of the total energy with respect to the FOD components can be computed, and these FOD ``forces'' can be used in a gradient-based minimization scheme to determine the optimal orbitals.  We note that the process of finding the $3N$ optimal FOD positions compares to finding the O($N^2$) elements of an optimal unitary transformation between canonical and localized orbitals in traditional PZ-SIC calculations.\cite{FERMI1,doi:10.1063/1.446959,doi:10.1063/1.448266}
 
 
 In LSIC,\cite{doi:10.1063/1.5129533} the exchange-correlation energy for a specific DFA 
 is written as

\begin{equation}\label{eq:LSIC}
    E_{XC}^{LSIC-DFA} = E_{XC}^{DFA}[\rho_{\uparrow}, \rho_{\downarrow}] - \sum\limits_{i,\sigma}^{occ} \{U^{LSIC}[\rho_{i\sigma}] + E_{XC}^{LSIC}[\rho_{i\sigma},0]\} \\
\end{equation} 
where  
\begin{equation}
    U^{LSIC} [\rho_{i\sigma}] = \frac{1}{2}\int d\vec{r} \{z_\sigma(\vec{r})\}^k \rho_{i\sigma(\vec{r})} \int d\vec{r'} \frac{\rho_{i\sigma}(\vec{r^\prime})}{|\vec{r}-\vec{r^\prime}|},\\
\end{equation}
    
\begin{equation}
     E_{XC}^{LSIC} [\rho_{i\sigma},0] = \int d\vec{r} \{z_\sigma(\vec{r})\}^k \rho_{i\sigma(\vec{r})}   \epsilon_{XC}^{DFA} ([\rho_{i\sigma},0],\vec{r}).
\end{equation}
Here the scaling factor
$z_{\sigma}(\vec{r})=\frac{\tau_\sigma^W(\vec{r})}{\tau_\sigma(\vec{r})}$ lies between 0 and 1 and indicates the nature of the charge density at $\vec{r}$.  
$z_\sigma=1$ for a density corresponding to a single electron orbital and 
$z_\sigma=0$ for a uniform density. Scaling the self-interaction correction terms with $z_\sigma$ thus retains the full correction 
for one-electron densities, making the theory exact in that limit, and eliminates the correction in the limit of a uniform density where 
$E_{XC}^{DFA}$ is exact.   As mentioned by Zope \textit{et al.}\cite{doi:10.1063/1.5129533} any iso-orbital indicator that can identity the single-electron region can be used 
in the LSIC method.
 We will refer this LSIC method as LSIC($z_\sigma$) to indicate use of LSIC method with kinetic energy density ratio as iso-orbital 
 indicator.
Using the pointwise LSIC correction results in significant improvement in electronic properties over SIC-LDA, particularly in situations where atoms are near the equilibrium positions and PZ-SIC performs poorly. 
 
 LSIC earlier was employed in a non-variational manner on self-consistent PZ-SIC density.\cite{doi:10.1063/1.5129533} For calculation of polarizabilities it is necessary to capture the density variation in response to an applied electric field. We used a quasi-self-consistent approach (like in SOSIC\cite{doi:10.1063/5.0004738}) 
 using the following Hamiltonian:
 \begin{align} \label{eq:ham}
     H_j = - \frac{1}{2} \nabla^2 + v(\vec{r}) + \int \frac{\rho(\vec{r'})}{|\vec{r}-\vec{r'}|} d\vec{r'} + v_{XC}^{DFA}([\rho],\vec{r}) 
     -z_{\sigma}(\vec{r}) \left( 
     \int \frac{\rho_j(\vec{r'})}{|\vec{r}-\vec{r'}|} d\vec{r'}
     + v_{XC}^{DFA}([\rho_j],\vec{r}) \right) 
 \end{align}
 
This Hamiltonian ignores the density dependence of scaling factor $z_{\sigma}$ and 
 can be viewed analogous to the model  exchange (-correlation) potentials designed to provide accurate excitation energies or 
 polarizabilities.\cite{van1994exchange,van2001influence,gruning2002required,banerjee2007time,banerjee2008ab,schipper2000molecular}
 %
 %
 %
 Although this approach ignores the full variation of the LSIC term, we  expect most effects of 
 self-consistency (as in orbital-dependent-SIC\cite{doi:10.1063/5.0004738}) are captured using the scaled SIC potentials.  

 The static dipole polarizability tensor $\alpha_{ij}$ is defined from change in the total energy $E$ of a 
 system under an applied static electric field $\vec{F}$ as
    \begin{eqnarray}
        E(\vec{F}) & = & E(0)+ \,\sum_i\frac{\partial E}{\partial F_i}F_i \,+ \,\frac{1}{2}\sum_{ij} \frac{\partial^2E}{\partial F_i \partial F_j}F_iF_j \,+ \cdots\\
          & = & E(0)- \,\sum_i \mu_iF_i- \,\frac{1}{2}\sum_{ij}\alpha_{ij} F_i F_j \,+ \cdots,
     \end{eqnarray}
 where $E(0)$ is the energy at zero electric field and $\mu_i$ and $F_i$ are the $i^{th}$ component of the dipole moment and the applied electric field. 
%
%
  We compute the components of the polarizability tensor using the finite-difference formula as follows
  \begin{equation}
      \alpha_{ij}=\frac{\partial \mu_i}{\partial F_j}\bigg\rvert_{F_j=0} =\lim_{F_j \to 0} \frac{\mu_i( F_j)-\mu_i(- F_j)}{2 F_j} .
\end{equation}
 The finite difference calculation of polarizability using the dipole has been proven to be more accurate than the sum-over-states perturbation method.\cite{mol-pol}
 Electric fields of strength 0.005 a.u. were applied in $\pm x$, $\pm y$, and $\pm z $ directions to determine the components of the polarizability tensor for each cluster. The average polarizability $\alpha_{avg}$ is calculated from the trace of the polarizability tensor as 
 \begin{equation}
      \alpha_{avg}=\frac{1}{3}\sum_{i}^3\alpha_{ii} .
\end{equation}
 The anisotropy $\beta$ of the polarizability tensor is calculated as 
 \begin{equation} 
     \beta^2=\frac{1}{2}\left[(\alpha_{xx}-\alpha_{yy})^2+(\alpha_{xx}-\alpha_{zz})^2+(\alpha_{yy}-\alpha_{zz})^2 + 6(\alpha_{xy}^2 + \alpha_{xz}^2 + \alpha_{yz}^2 )\right]\\ .
 \end{equation}
 
 
 The standard Gaussian orbital basis sets used in the NRLMOL and FLO-SIC codes are optimized for the PBE functional\cite{PBE} and are of quadruple zeta quality.\cite{Freiberg}  For H (O), the standard basis set includes 4 (5) s-type, 3 (4) p-type, and 1 (3) (Cartesian) d-type orbitals, contracted from a set of 7 (14) single Gaussian orbitals (SGO). For polarizability calculations, we used an additional SGO polarization function for each angular moment type. 

 We used non-empirical functionals from the lowest three rungs of Jacob's ladder to compute 
 polarizabilities with and without SIC: LDA as parameterized in PW92,\cite{PhysRevB.45.13244} PBE,\cite{PBE} and SCAN.\cite{SCAN} Calculations with  SCAN  require a particularly fine grid for numerical accuracy.\cite{yamamoto2019fermi} 
  For example, for water hexamer isomers, the SCAN grid contains roughly twice the number of points
 than the grid needed for LDA and PBE.

  \begin{figure}
     \centering
     \includegraphics[width=\textwidth]{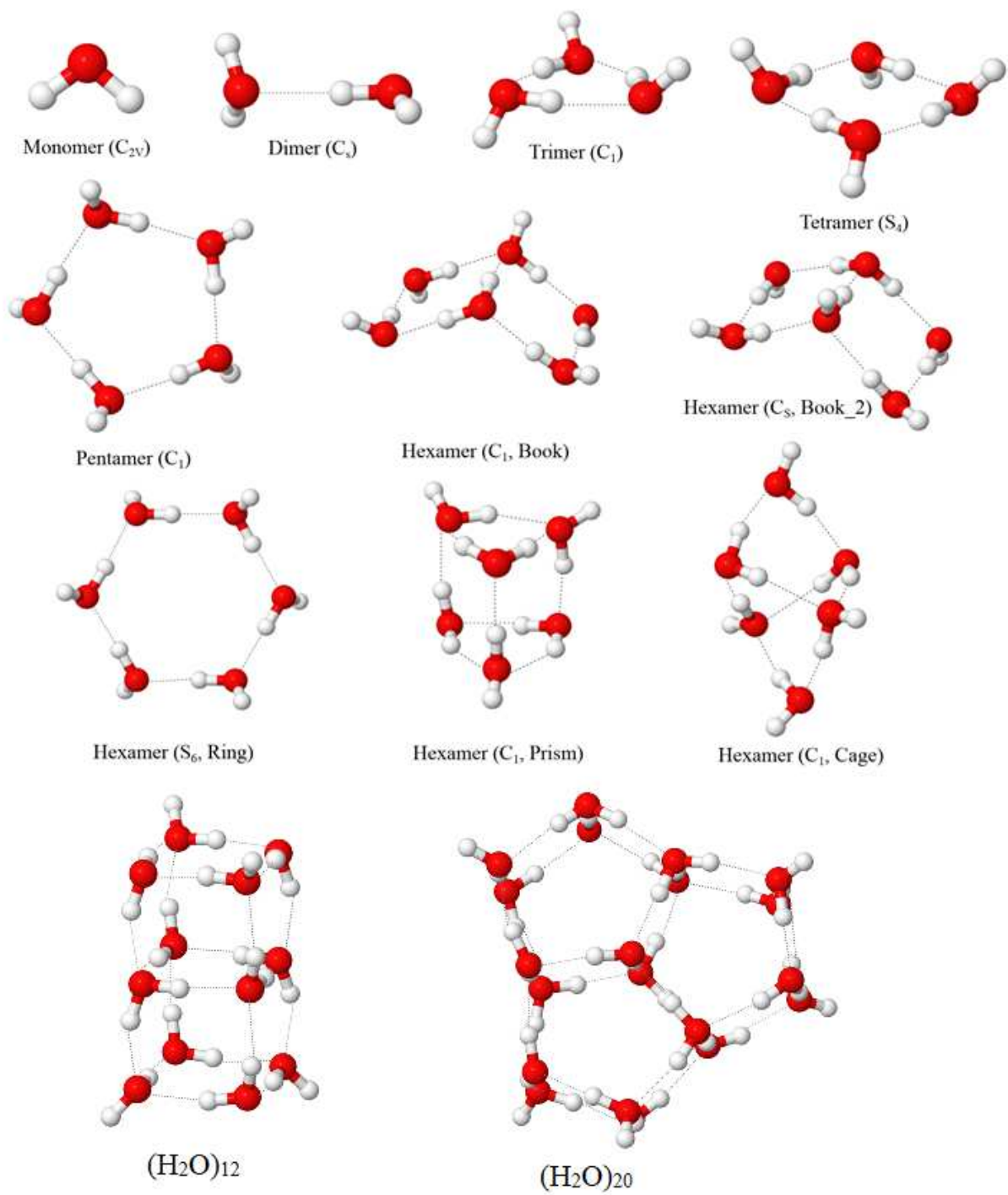} 
     \caption{The structures of the water clusters of sizes $n=1-6,12$, and $20$.}
     \label{fig:geometry}
 \end{figure}


 \section{Results and Discussion} \label{sec:results}
 
  We used the geometries of the water clusters optimized at the CCSD(T) level of theory by Miliordos \textit{et al.}.\cite{doi:10.1063/1.4820448} 
  This set of clusters contains one isomer each for the monomer to pentamer, and five isomers
  for the water hexamer. Two larger clusters of twelve and twenty water molecules optimized at the MP2/aug-cc-PVTZ level\cite{doi:10.1063/1.5128378} are also included in this set. The cluster structures are shown in Fig. \ref{fig:geometry}. 
  The hexamers are labeled according to shape as  prism (p), ring (r), book1 (b1), book2 (b2), and cage (c). 
 The binding energies of the water hexamers found here follow the
 ordering reported in Ref. \cite{FLOSIC_WATER_PNAS}.  The stability ordering using LDA, SCAN, FLO-SIC-LDA, and FLO-SIC-SCAN from the most stable to the least is p$<$c$<$b1$<$b2$<$r whereas using PBE and 
 FLO-SIC-PBE the ordering is b1$<$r$<$c$<$p$<$b2. 
 
 We optimized the positions of the FODs independently for FLO-SIC-LDA, FLO-SIC-PBE, and FLO-SIC-SCAN calculations.  In addition, because the electron distribution may change with application of an external field, we re-optimized the FOD positions when an external field was applied, using the zero-field positions as starting points.  
 We find that the FOD positions generally relax in the direction opposite to the applied electric field. 
 This shift in position is small, for example, when a static electric field of strength 0.005 a.u. is applied where the largest shift in position of FOD is 0.2 \AA.
 Re-optimizing the FOD positions has a relatively small effect on the value of the calculated polarizability.  For example, in FLO-SIC-PBE, the polarizability of the monomer changes from 8.34 Bohr$^3$ using the zero-field  FODs for all calculations, to 8.55 Bohr$^3$ using FODs that are optimized for each applied field. For hexamers, the average change in polarizability arising from FOD optimization ranges from 1.8 to 2.2\% with LDA and PBE, respectively. As can be seen below, 
 this difference is small compared to the differences between DFA and FLO-SIC-DFA results.
 


                   
  
\begin{table*}
\caption{The dipole moments of water clusters molecules (in Debye) with DFA, FLO-SIC, and LSIC($z_\sigma$). The MAEs are with respect to MP2.}
\label{table_dipole}
\begin{tabular*}{0.96\textwidth}{@{\extracolsep{\fill}}cccccccccc}
\colrule
Water  &\multicolumn{3}{c}{DFA} & \multicolumn{3}{c}{FLO-SIC}& \multicolumn{2}{c}{LSIC($z_\sigma$)}& MP2\footnote{Reference [\onlinecite{gregory1997water}]} \\\cline{2-4} \cline{5-7}  \cline{8-9} 
                   
  cluster & LDA & PBE & SCAN & LDA & PBE & SCAN & LDA & PBE & \\  \hline
  1&  1.90& 1.84&	1.88&	2.00& 1.94&		1.97&	1.87 &1.84&1.868\\     
  2&  2.71& 2.62&	2.63&	2.74& 2.68&		2.70&	2.61 &2.58&2.683 \\
  3&  1.08& 1.05&	1.07&	1.15& 1.12&		1.13&	1.09 &1.07&1.071 \\
  4&  0.00&	0.00&	0.00&	0.00& 0.00&	    0.00&	0.00 &0.00& 0.00\\
  5&  1.01& 0.98&	1.00&	1.07& 1.04&		1.05&	1.01 &1.00& 0.927\\
  6p& 2.68& 2.59& 	2.62&	2.77& 2.69&		2.72&	2.61 &2.57&2.701\\
  6c& 2.08& 2.01&	2.03&	2.13& 2.07&		2.10&	2.00 &1.97&1.904\\
  6b1&2.55& 2.46&	2.49&	2.63& 2.55&		2.57&	2.46 &2.42& --\\
  6b2&2.59& 2.50&	2.53&	2.69& 2.60&		2.62&	2.51 &2.46& --\\
  6r& 0.00& 0.00&	0.00&	0.00& 0.00&	    0.00&	0.00 &0.00&0.00\\
  12 & 0.00& 0.00 & 0.00 &0.00 & 0.00 & -- & 0.00 & -- & --\\
  20 & 0.28 & 0.28 & 0.29 & 0.30 &-- &-- &0.30 &-- &-- \\
  \colrule
 MAE& 0.04& 0.05& 0.04& 0.09  &0.05  &0.06 &0.05&0.05\\
 \colrule
      \end{tabular*}
  \end{table*}

                   
 
 The zero-field static dipole moments calculated for the clusters are presented in Table \ref{table_dipole}. 
 Gregory \textit{et al.} calculated 
 the dipole moments of small water clusters up to  $n=6$ using the MP2 theory and aug-cc-VDZ basis sets.\cite{gregory1997water} These are included for reference in Table \ref{table_dipole}.  We emphasize that although there are small differences in the geometries of the reference set and ours, these comparisons can still show the trend of  dipole moments with different DFAs with FLO-SIC and LSIC($z_\sigma$).  
  The application of FLO-SIC increases the size of non-zero dipole moments in the water clusters 
  compared to DFA calculations.  Similar observations were made in an earlier FLO-SIC application to 
 a diverse set of molecules.\cite{FLOSIC_dipole}  This effect can be understood considering the effect that SIC has on the potential seen by the electrons.  
 By removing the interaction of an electron with itself, SIC makes the potential more attractive.  For example, for a neutral atom or molecule, the DFA potential decays exponentially to zero in the asymptotic region.  SIC corrects this unphysical behavior, restoring the correct $-1/r$ form.  In a heteronuclear molecule, this tends to stabilize the anionic part relative to the cationic part, giving rise to larger dipoles.  In Table \ref{table_dipole}, the FLO-SIC-DFA dipoles are between $0.03$ and $0.10$ D larger than the corresponding DFA values in all cases except for the tetramer, the ring hexamer, and the 12-molecule water that have exactly zero dipole moments due to their symmetry. 
  %
  The directions of the dipoles of the water clusters remained nearly same in all 
  clusters with DFA and FLO-SIC-DFA. 

  The water cluster dipole moments were also calculated using the recently developed 
  LSIC($z_\sigma$) approach with LDA and PBE\cite{doi:10.1063/1.5129533} functionals. 
   For the monomer the LDA/FLO-SIC-LDA dipole moments are 1.90 D and 
   2.00 D, respectively, while LSIC($z_\sigma$)-LDA results in a value of 1.87 D, much  closer to the experimental  value of 1.855 D.\cite{Lovas1978}
   The dipole moments for the dimer with PBE and SCAN are close to the experimental 
   value of 2.643 D\cite{Dyke1977} which increase to 2.68 D and 2.70 D with 
   SIC. In this case LSIC($z_\sigma$)-LDA predicts a somewhat smaller value of
   2.61 D, but much closer to the experimental value than the FLO-SIC-LDA value (2.74 D).
  On the whole, the LSIC($z_\sigma$)-LDA dipole values have a MAE of 0.05 D, a significant improvement over FLO-SIC-LDA (0.09 D), but similar to the other methods.
  Surprisingly, the LSIC($z_\sigma$)-LDA values are generally smaller than the LDA values.  A possible explanation for this is that the density near the H atoms in a water cluster is dominated by the H 1s orbital and is thus one-electron-like.  Thus, full SIC is expected in the region near the H atoms, whereas a reduced SIC is expected near the O atoms, effectively lowering the potential seen by an electron near the H atoms relative to that near O.  The net effect of this may be to shift electron charge toward the H atoms, resulting in smaller dipoles.  This effect is currently under further investigation.

\begin{table*}[!htbp] 
      \caption{The average polarizability of water clusters per water molecule (in Bohr$^3$) obtained with DFA, FLO-SIC, and LSIC($z_\sigma$). The MAEs are with respect to the two sets of CCSD reference values.}
      \label{table_pol}
\begin{tabular*}{0.96\textwidth}{@{\extracolsep{\fill}}ccccccccccc}
\colrule
Water  &\multicolumn{3}{c}{DFA} & \multicolumn{3}{c}{FLO-SIC}& \multicolumn{2}{c}{LSIC($z_\sigma$)}& CCSD$^a$ & CCSD$^b$


\\\cline{2-4} \cline{5-7}  \cline{8-9} 
                   
   cluster & LDA & PBE & SCAN & LDA & PBE & SCAN & LDA & PBE & & \\  \hline
  1&   10.63 & 10.58 & 10.04  & 8.43    & 8.55& 8.69& 9.27& 9.28 & 9.26 & -- \\ 
  2&   10.80 & 10.74 & 10.17  & 8.40    & 8.70& 8.75& 9.31& 9.37 & 9.55 & 9.30\\
  3&   10.84 & 10.76& 10.18   & 8.48    & 8.70& 8.68& 9.32& 9.35 & 9.65 & 9.37\\
  4&   10.86 & 10.79& 10.24   &  8.59   & 8.84& 8.81& 9.40& 9.44 & 9.76 & 9.48\\
  5&   10.92 & 10.85& 10.28   &  8.62   & 8.88& 8.82& 9.43& 9.49 & 9.82 & 9.53\\
  6p&  10.72 & 10.61& 10.21   &  8.46   & 8.70& 8.68& 9.22& 9.24 & 9.61 & 9.34\\
  6c&  10.78 & 10.68& 10.13   &  8.49   & 8.77& 8.69& 9.29& 9.32 & 9.68 & 9.40\\
  6b1& 10.86 & 10.81& 10.19   &  8.59   & 8.87& 8.80& 9.41& 9.44 & 9.79 & 9.50\\
  6b2& 10.89 & 10.80& 10.25   &  8.59   & 8.86& 8.77& 9.38& 9.44 & --    & --\\
  6r&  10.96 & 10.89& 10.30   &  8.65   & 8.91& 8.79& 9.46& 9.51 & 9.86    & 9.56\\
  12 & 10.60 & 10.47& 9.99 &  8.32   &  --     & -- &  9.07     & --   & -- & -- \\
  20 & 10.66 &--&--& 8.18 &--&--& 9.12 &--&--&--\\
  \colrule
   MAE$^a$ & 1.15 & 1.08 & 0.53 & 1.14 & 0.90 & 0.92 & 0.32 & 0.29 && \\
   MAE$^b$ & 1.41 & 1.33 & 0.78 & 0.90 & 0.64 & 0.68 & 0.08 & 0.06 &&\\
   \colrule
\multicolumn{10}{l}{\small{$^a$ Reference [\onlinecite{Hammond}]}}\\
\multicolumn{10}{l}{\small{$^b$ Reference [\onlinecite{Alipour2013}]}}\\
      \end{tabular*}      
  \end{table*}

  In Table \ref{table_pol}, we list isotropic dipole polarizabilities per water molecule for all the clusters,
 calculated using the three DFAs, FLO-SIC-DFAs, LSIC($z_\sigma$)-LDA, and LSIC($z_\sigma$)-PBE methods. The polarizabilities are calculated using the finite difference method which requires application of an electric field along different directions. 
 Hammond \textit{et al.} carried out calculations of the dipole polarizability of water clusters up to $n = 12$ at the coupled-cluster level of theory\cite{Hammond} 
 using MP2 geometries
\cite{Xantheas_MP2_1, Xantheas_MP2_2} using linear response theory.
 Alipour and co-workers also compared water cluster polarizabilities obtained with coupled-cluster,  hybrid and double hybrid DFT methods\cite{Alipour2013, Alipour2017} on geometries optimized at the CCSD/aug-cc-pV level, but employing the finite difference approach.
 There are small, but systematic differences between the reference values, with those from Ref. \cite{Alipour2013} approximately 0.25 Bohr$^3$ per molecule smaller for each cluster size.  Due to these differences, we computed mean absolute errors with respect to both sets of reference values.

 %
  
   By removing self-interaction effects, the FLO-SIC-DFA methods give smaller polarizabilities. For the water monomer, for example, 
 polarizability values with DFAs range from $10.04-10.63$ 
 Bohr$^3$, which reduce to $8.43-8.69$ Bohr$^3$ with FLO-SIC
 The experimental isotropic 
  polarizability of the water monomer was reported as 9.63 Bohr$^3$ 
  by Zeiss and Meath from refractive index data and dispersion measurements\cite{Zeiss1975} 
  and 9.92 Bohr$^3$ by Murphy based on rovibrational Raman scattering measurements.\cite{Murphy1977} The present calculations are restricted 
  to only the static dipole polarizability without considering vibrational contributions that would increase the effective polarizability.\cite{Pederson2005} 
   
  A similar trend is seen for the larger water clusters. The DFA polarizabilities overestimate the reference CCSD values, while the FLO-SIC-DFA polarizabilities underestimate them. 
  The MAEs are similar for  LDA and PBE at the DFA level and lower for SCAN.
  Interestingly,  FLO-SIC-PBE and FLO-SIC-SCAN have comparable MAEs that are significantly smaller than for FLO-SIC-LDA. The SCAN functional is self-correlation free and 
  the SCAN results are improved only slightly with FLO-SIC.

\begin{figure}[!ht]
  \centering

    \includegraphics[width=0.9\textwidth]{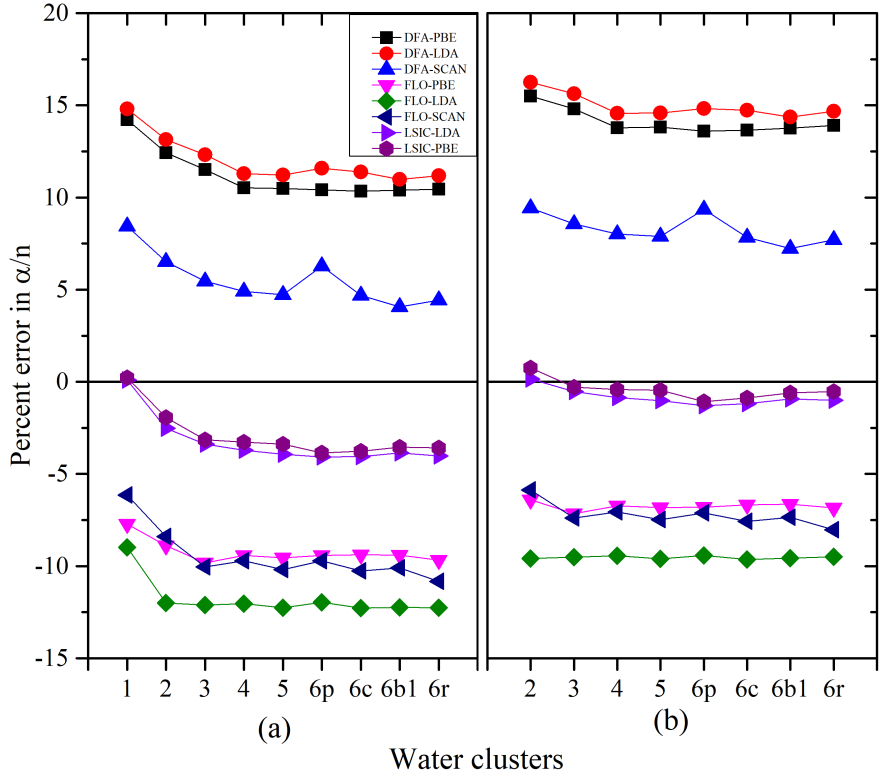}
  \caption{Percentage errors of calculated polarizability per water molecule with respect to CCSD values reported in (a) Ref. \cite {Hammond} and (b) Ref. \cite{Alipour2013}. The label FLO in the plot indicates FLO-SIC, and LSIC indicates LSIC($z_\sigma$).}
  \label{fig:err}
\end{figure}

In Fig.~\ref{fig:err} the percentage deviations of the calculated polarizabilities per molecule  from the corresponding CCSD values of Ref. \cite{Hammond} and \cite{Alipour2013} are plotted for all three DFAs, associated FLO-SIC-DFAs.
It is interesting that the errors for LDA and PBE fall close together, lying between $10-14$\%,
 whereas the errors for SCAN range from $4-8$ \%.  For the FLO-SIC-DFA results, FLO-SIC-PBE and FLO-SIC-SCAN errors are very similar, underestimating CCSD by $5-10$\%, while FLO-SIC-LDA underestimates CCSD by $8-13$\%. 
 




LSIC($z_\sigma$) was applied with LDA and PBE. Although  gauge-consistency of the energy density in LSIC($z_\sigma$) is rigorously satisfied only for LDA, it was shown by Bhattarai \textit{et al.}\cite{sdSIC} that the gauge violation does not severely affect the LSIC($z_\sigma$)-PBE results.  On the other hand, LSIC($z_\sigma$)-SCAN was found to be strongly affected\cite{sdSIC} so we have not done LSIC($z_\sigma$)-SCAN calculations here. 
  For polarizabilities of water clusters, we find that LSIC($z_\sigma$) shrinks 
  the excessive correction of PZ-SIC, 
  resulting in values that lie between the DFA and FLO-SIC-DFA results in all cases, bringing them to excellent agreement with the corresponding CCSD results. The  MAE for LSIC($z_\sigma$)-LDA and LSIC($z_\sigma$)-PBE are 0.32 and 0.29 Bohr$^3$ with respect to Ref. \cite{Hammond} and 0.08 and 0.06 Bohr$^3$ with Ref. \cite{Alipour2013}. 
  LSIC($z_\sigma$) clearly performs best among the methods employed here. 
 
 The larger $n =$ 12 and 20 clusters are relatively more compact and three dimensional than the smaller clusters. They have relatively more open interior space and therefore likely to exhibit larger screening effects. 
   It is of interest to gain some insight into whether the same relationship between DFA, PZ-SIC, and LSIC($z_\sigma$) polarizabilities continues as these effects begin to become important. Going from $n =$ 6 to 12 to 20, we see a significant drop in the polarizability per molecule at the DFA, as well as the PZ-SIC and LSIC($z_\sigma$) levels due to screening.  
A similar drop was noted by Hammond \textit{et al.} between $n=6$ and $12$ clusters\cite{Hammond}. 
 It also appears that the same relationship for the $\alpha/n$ values between different methods  continues for the larger clusters:  DFA $>$ LSIC($z_\sigma$) $>$ FLO-SIC.  
  
  
        We have also calculated the anisotropic polarizability for the water 
  clusters and the values for all the functionals are presented in Table \ref{table_ani}. 
  CCSD values computed with the aug-cc-pVDZ basis\cite{Hammond,Alipour2013} are 
  also shown.  
  The anisotropy of polarizability of the larger clusters show an opposite trend compared to the monomer. For the monomer the DFA values are much smaller compared to CCSD values and FLO-SIC values are too large. LSIC($z_\sigma$) brings the values down closer to the reference values.
  For the larger clusters on the other hand DFA values are too large compared to CCSD values.
  The FLO-SIC values are lowered with lower MAEs. Unlike the isotropic polarizability, FLO-SIC does 
  not degrade the performance of SCAN for the anisotropic component. LSIC($z_\sigma$) brings it even closer to the reference values. We find that the MAEs for LSIC($z_\sigma$)-LDA and LSIC($z_\sigma$)-PBE are 0.52 and 0.37 Bohr$^3$ for   the clusters up to the hexamers. At the DFA level, SCAN has the lowest MAE but at the FLO-SIC and LSIC($z_\sigma$) level PBE functional provides the best description.
  
  
       
\begin{table*}[ht] 
\caption{ { The anisotropic polarizability of water clusters (in Bohr$^3$) using the DFA, FLO-SIC, and LSIC($z_\sigma$) approaches. The CCSD reference values with the 
aug-cc-pVDZ (DZ) basis.  
The MAEs are with respect to CCSD.}}
\label{table_ani}
\begin{tabular*}{1.00\textwidth}{@{\extracolsep{\fill}}cccccccccccc}
\colrule
Water  &\multicolumn{3}{c}{DFA} & \multicolumn{3}{c}{FLO-SIC} & 
\multicolumn{2}{c}{LSIC($z_\sigma$)} & CCSD/DZ & CCSD/TZ$^a$ & CCSD/QZ$^a$\\\cline{2-4} \cline{5-7}  \cline{8-9} 
                   
 cluster  & LDA & PBE & SCAN & LDA  & PBE  & SCAN & LDA & PBE       &         &          &.       \\  \hline
  1&0.23&  0.24&  0.51& 1.56 & 1.62 & 2.10 & 1.21&1.26   & 1.03$^a$    &0.71      &0.61    \\
  2&4.20&  4.11&  3.74& 3.42 & 3.59 & 3.99 & 3.57&3.54   &3.47$^a$     &3.24      &2.83     \\
  3&6.47&  6.43&  5.87& 5.53 & 5.78 & 5.27 & 5.66&5.72   &5.28$^a$     &5.21      &4.87     \\
  4&9.66&  9.55&  8.61& 6.72 & 7.08 & 7.08 & 7.57&7.42   &7.46$^a$     &7.51      & --\\
  5&12.50& 12.37&11.10& 8.38 & 9.28 & 8.27 & 9.76&9.55   &9.15$^b$     &--         &--\\
 6p&5.30&  5.51&  5.32& 6.00 & 5.95 & 6.25 & 5.92&5.86  &5.21$^b$     &-- &--\\
 6c&9.57&  9.49&  8.844& 8.81 & 8.64 & 8.70 & 8.98&8.75 &7.70$^b$     &--&--\\
6b1&11.84&11.40& 10.31 & 8.69 & 8.77 & 8.74 & 9.55&9.05 &8.80$^b$&--&--\\
6b2&10.53&10.48&  9.52& 6.89 & 8.50  & 8.20 & 8.42&8.55 &    --   &--&--\\
6r &14.99&14.80&  12.97& 9.66 & 10.25 & 9.54 & 11.07&10.76&10.55$^b$ &-- &--\\
\colrule
 MAE &  1.97 &	1.87  & 1.07 &	0.58 & 	0.41 & 0.66 &0.52 &0.37 &&\\
\colrule
\multicolumn{10}{l}{\small{$^a$ Reference [\onlinecite{Hammond}]}}\\
\multicolumn{10}{l}{\small{$^b$ Reference [\onlinecite{Alipour2013}]}}\\
\end{tabular*}
  \end{table*}

 The calculated dipole moments and the polarizability of the water clusters demonstrate that the local scaling of SIC performs best among the methods employed here. As mentioned in the introduction, the polarizability depends mostly on the valence electron density which is determined by the potential in the asymptotic region. To understand further how FLO-SIC and LSIC($z_\sigma$) affect the valence electron, we compare the negative of the eigenvalues of the highest occupied orbitals with vertical ionization potentials (vIPs) calculated with CCSD(T)/aug-cc-PVTZ levels on geometries optimized at the CCSD/aug-cc-PVDZ level.\cite{doi:10.1063/1.4730301} 
 vIP is the transition energy between the ground state energy of a neutral system and the lowest energy of the cation that has the same geometry as the neutral system. The adiabatic IP, on the other hand, is the energy difference between the ground states of a neutral system and the cation where the cation geometry is relaxed. 
 Segarra-Mart\'i \textit{et al.} have shown that the difference between vIP and adiabatic IP for water clusters calculated with CCSD(T) is large ($\sim$2 eV).\cite{doi:10.1063/1.4730301} Since the HOMO eigenvalues best represent the vIP without any structural relaxation, we restrict our comparison to theoretical vIP values only. 
 These comparisons of DFA, FLO-SIC-DFA, and LSIC($z_\sigma$)-DFA are presented in Table \ref{Table_ip}. The HOMO eigenvalues at the DFA level is too high due to the incorrect asymptotic decay of the potential. FLO-SIC (PZ-SIC) corrects for that, however, it becomes too deep. The MAEs for the HOMO eigenvalues are 2.37, 1.82, and 1.97 eV for  FLO-SIC-LDA, FLO-SIC-PBE, and FLO-SIC-SCAN respectively. With the LSIC($z_\sigma$) approach, the HOMO eigenvalues show excellent agreement with CCSD(T) results with  MAEs of 0.40 and 0.06 eV with LDA and PBE functionals.  For the sake of comparison, the self-consistent GW values from Ref. \cite{doi:10.1063/1.4940139} on the same CCSD/aug-cc-PVDZ geometries are also included. The MAE of the GW@PBE0 with respect to the CCSD(T) results is 0.35 eV, which is comparable to the LSIC($z_\sigma$)-LDA MAE. 
 The HOMO eigenvalues show that the LSIC($z_\sigma$)-PBE describes the valence region most accurately among the methods employed here. These results are in accord with the results for polarizability where the MAE for LSIC($z_\sigma$)-PBE is found to be the smallest and the LSIC($z_\sigma$)-LDA is the second best approximation. Moreover, LSIC($z_\sigma$)-PBE values are in better agreement with CCSD(T) than GW@PBE values (MAE, 0.20 eV). This shows that the local scaling of SIC can achieve similar accuracy as more sophisticated methods.  The region around the H atoms are the one-electron like regions where full SIC is applied in the LSIC($z_\sigma$) method. This is also the valence region where the SIC correction is required for good HOMO eigenvalues. Alternatively, the scaling down of SIC in the many electron region near the oxygen is necessary for a good description of polarizability and vertical ionization properties. The success of LSIC($z_\sigma$) also indicates that the selective orbital based SIC (SOSIC) method\cite{doi:10.1063/5.0004738} where the SIC is applied to the selected set of orbitals (for example, HOMO) is also likely to be successful for water clusters. More applications of LSIC are necessary to determine its effectiveness for various systems and properties.

 \begin{table*}[ht] 
\caption{ { The HOMO eigenvalues of water clusters (in eV) using the DFA, FLO-SIC, and LSIC($z_\sigma$) approaches.
The MAEs are with respect to CCSD(T).}}
      \label{Table_ip}
%
%
%
%

\begin{tabular*}{0.96\textwidth}{@{\extracolsep{\fill}}cccccccccccc}
\colrule
Water  &\multicolumn{3}{c}{DFA} & \multicolumn{3}{c}{FLO-SIC} &  \multicolumn{2}{c}{LSIC($z_\sigma$)}& GW@PBE$^a$ & GW@PBE0$^b$ & CCSD(T)
\\\cline{2-4} \cline{5-7}  \cline{8-9} 
cluster& LDA & PBE & SCAN & LDA  & PBE  & SCAN & LDA & PBE & & & \\ \hline
1&	7.34&	7.20&	7.58&	14.72&	14.23&	14.26&	12.97&	12.64& 12.88&    --& 12.653$^c$\\
2&	6.64&	6.49&	6.88&	13.98&	13.44&	13.59&	12.13&  11.79& 12.03& 12.15& 11.79$^d$\\
3&	7.34&	7.15&	7.54&	14.62&	14.06&	14.23&	12.66&	12.29& 12.45& 12.60& 12.27$^d$\\
4&	7.44&	7.25&	7.63&	14.68&	14.12&	14.28&	12.69&  12.32& 12.49& 12.63& 12.27$^d$\\
5&	7.28&	7.09&	7.48&	14.53&	13.96&	14.14&	12.54&	12.17& 12.32& 12.47& 12.10$^d$\\
6p&	6.99&	6.80&	7.18&	14.22&	13.66&	13.82&	12.18&	11.81& 11.88& 12.04& 11.65$^d$\\
6c&	7.10&	6.91&	7.30&	14.34&	13.78&	13.94&	12.30&	11.93& 12.04& 12.18& 11.99$^d$\\
6b1& 7.03&	6.83&	7.22&	14.22&	13.62&	13.82&	12.17&	11.79& 11.91& 12.07& 11.69$^d$\\
6b2& 7.00&	6.81&	7.20&	14.20&	13.63&	13.30&	12.13&	11.76&    --&    --&    --\\
6r&	 7.35&	7.15&	7.54&	14.58&	14.02&	14.19&	12.56&	12.19& 12.40& 12.54& 12.14$^d$\\
12&  7.46&  7.25&   7.63&   14.63&     --&     --&  12.51&     --&    --&    --&    --\\
20& 7.37&   7.16&   7.55&   14.56&     --&     --&  12.40&     --&    --&    --&    --\\
\colrule
MAE& 4.89&	5.08&	4.69&	 2.37&	 1.82&	 1.97&   0.40&   0.06&  0.20&  0.35&    --\\
\colrule
\multicolumn{11}{l}{\small{$^a$GW@PBE with aug-cc-pVTZ basis from reference [\onlinecite{doi:10.1063/1.4940139}]}}\\
\multicolumn{11}{l}{\small{$^b$GW@PBE0 with aug-cc-pVQZ and Weigend auxiliary basis from reference [\onlinecite{doi:10.1063/1.4940139}]}}\\
\multicolumn{11}{l}{\small{$^c$Reference [\onlinecite{doi:10.1063/1.4940139}]}}\\
\multicolumn{11}{l}{\small{$^d$Reference [\onlinecite{doi:10.1063/1.4730301}]}}\\
\end{tabular*}
\end{table*}

  \section{Summary}\label{sec:summary}
  
 The effect of SIEs in the DFAs, viz. LDA, PBE-GGA, and SCAN meta-GGA functionals
 on the dipoles and static dipole polarizabilities of water clusters was studied using the FLO-SIC method and LSIC($z_\sigma$)
 methods as implemented in the FLO-SIC code.  
 Our results show that poor description of the valence regions by DFAs  results in 
 overestimation of water cluster isotropic polarizabilities. The SCAN functional however performs relatively better than PBE or LDA with $40-50\%$ lower MAE. 
 Incorporation of SIC results in reducing the polarizabilities however the corrected  polarizability values are too small compared to the reference values. 
 Pointwise correction with the LSIC($z_\sigma$) method brings the isotropic polarizability values with LDA and PBE closer to the reference values. 
 Similar trends are also noticed for anisotropy in polarizability and dipole moments for which the LSIC($z_\sigma$) values are closer to the reference values compared to either DFA or FLO-SIC-DFA. The FLO-SIC (PZ-SIC) correction is on orbital by orbital basis that treats both one-electron region and many-electron regions equally. On the other hand, the LSIC($z_\sigma$) keeps the complete SIC in the one-electron like regions and scales down the correction in many-electron regions. This approach, when applied with LDA and PBE showed excellent agreement for polarizabilities with reference CCSD values. 
 We also show that the HOMO eigenvalues of the clusters with LSIC($z_\sigma$)-LDA and LSIC($z_\sigma$)-PBE are in excellent agreement with ionization energies.
 The results presented here, combined with earlier 
 applications on water cluster anions,\cite{C9CP06106A} show that SIC is important for describing the electronic properties of water.  
 This work also shows that LSIC($z_\sigma$) gives an excellent description of electronic properties of water cluster.
 More applications with 
 LSIC are necessary to determine its effectiveness for other types of systems and other properties. Variations of scaling of PZ-SIC \cite{doi:10.1063/5.0004738, sdSIC} also need to be tested to more clearly understand the failures of PZ-SIC, which can lead to development of better functionals.

\section*{Data Availability Statement}
The data that supports the findings of this study are available within the article.

\begin{acknowledgments}  

This work was supported by the US Department of Energy, Office of 
Science, Office of Basic Energy Sciences, as part of the 
Computational Chemical Sciences Program under Award No. 
DE-SC0018331. The work of T.B. and Y.Y. was supported in part by
the US Department of Energy, Office of Science, 
Office of Basic Energy Sciences, under Award No. DE-SC0002168.
Support for computational time at the Texas Advanced 
Computing Center through NSF Grant No. TG-DMR090071 
and at NERSC is gratefully acknowledged.
\end{acknowledgments} 
  
 \bibliography{water_pol_ref}
\end{document}